\documentclass[11pt,twoside]{article}

\usepackage{asp2006}
\usepackage{epsf}
\usepackage{graphicx}
\usepackage{lscape}

\begin{document}
\title{The Formation of Low-Mass Protostars and Proto-Brown Dwarfs}   
\author{Jochen Eisl\"offel$^1$ and J\"urgen Steinacker$^{2,3}$}   
\affil{$^1$ Th\"uringer Landessternwarte Tautenburg, Sternwarte 5, 
         D-07778 Tautenburg, Germany, jochen@tls-tautenburg.de\\
       $^2$Max-Planck-Institut f\"ur Astronomie,
            K\"onigstuhl 17, D-69117 Heidelberg, Germany, stein@mpia.de\\
       $^3$Astronomisches Rechen-Institut am Zentrum f\"ur 
            Astronomie Heidelberg, M\"onchhofstr. 12-14, 
            D-69120 Heidelberg, Germany}    

\begin{abstract} 
The formation of low-mass protostars and especially of brown dwarfs 
currently are ``hot topics'' in cool star research. 
The talks contributed to this splinter session discussed how low in mass and
how low in luminosity objects might exist, if these substellar objects show
evidence for circum(sub)stellar disks, and how the bottom of the mass function
in young clusters after the formation process looks like.
In a lively open discussion, a vast majority of the speakers and the audience
expressed why, given the available data, a stellar-like formation mechanism
down to the lowest masses should be preferred.
\end{abstract}


\section{Introduction}   

Star formation is one of the great four themes of 'Origins' studied in
astronomy today, and special attention is attributed to the formation of
low-mass stars like our own Sun. Low-mass stars are thought to form from the
collapse of a low-density interstellar molecular cloud, producing a
high-density core which evolves into a flattened proto-planetary disk through
which material is accreted onto the growing central object.

Large ground-based telescopes and currently active satellite observatories
like $HST$, $CHANDRA$, $XMM$, and most outstandingly $SPITZER$ are delivering 
a wealth of new details, partially forcing us to re-conceive our conceptions 
of star formation. At the same time, these new data are preparing the ground 
for $ALMA$ and $Herschel$, which will come online in the near future.

The advanced numerical simulations of the complex evolution of collapsing
low-mass cores are about to enter a new era with the explicit inclusion of
heating and cooling by radiative transfer and with multi-wavelength modeling
of high-resolution images.

However, despite all of these new high-resolution observations and simulations
of low-mass star forming regions, the main controlling agents of the early
phase star formation process remain highly debated. The main goal of this
splinter session was to highlight ongoing progress in tackling the controlling
physical processes of the formation of low-mass proto-stars and proto-brown 
dwarfs. All talks contributing to this splinter session came from the
observational side. This bias was not intended by the conveners, but possibly
may reflect the fact that most of the efforts in this field are currently
done on the observational side.

\section{Very low mass cores and the start of the star formation process}

It became very clear again in this splinter session that in spite of the 
suggested scenarios that explain the formation of brown dwarfs by ejection 
of the least massive bodies from very young multiple systems or small-n
clusters \citep{reipurth01, bate03}, observers in general prefer the model of
star-like formation of substellar objects, i.e. via direct fragmentation of
dense cores \citep{boss01, padoan02, with06}.
A large part of this session was therefore dedicated to the question of how 
far down in the mass spectrum low-mass cores and low-luminosity 
proto-substellar objects would exist, and in what numbers.

Jane Greaves reported on a new survey for low-mass cores carried out with the
$SCUBA$ submillimetre camera at the $JCMT$, which mapped the thermal dust 
emission
in the Ophiucus B and D clouds. Since this emission is optically thin, core
dust masses can be derived directly. 

In these observations, blobs are found well into the planetary mass regime,
with core masses of 10 M$_{jup}$ and less. One 9 Jupiter-mass object even 
shows a bipolar outflow. The number counts of low-mass cores suggest a simple
extension of the clump mass function from the stellar regime. It can be
described by a simple power law of the form  ${ dN / dlog(M) \propto 
M^{-0.5} }$ in the range 0.003 to 10 M$_{\odot}$. Mapping deeper and deeper, 
new faint high density blobs appear and the bottom of the core mass function 
is not yet reached at 3 M$_{jup}$.

Neil Evans presented results from the Core-to-Disk (c2d) legacy survey with
the $SPITZER$ space telescope. Using $IRAC$ mid-infrared data and deep 
optical/near-infrared images, a number of very low mass brown dwarfs with 
disks have been found in regions of star formation, and he addressed the 
question of how low in luminosity they go.
The optical/near-infrared data have found the candidate young brown
dwarfs, while the $SPITZER$ data have identified those with disks from their 
infrared excess emission. Interestingly, this excess emission seems to be
constant over a range of luminosities of the central object. Masses at or
below 13 Jupiter masses have been inferred for the lowest-mass central 
objects. These are found both near cluster formation regions and far from 
rich clusters, indicating that ejection from clusters cannot explain the 
formation of all brown dwarfs. 

In addition, a wide binary brown dwarf has been found in Ophiucus, 
which also argues against the cluster ejection scenario \citep{all05}. 

These results beg the question: if brown dwarfs form as stars do, where are
the earlier, embedded phases. Some embedded very low luminosity objects may be
the precursors of brown dwarfs with disks. A number of these have been found by
the c2d project in what had been thought to be starless cores. 
A prominent case is the L1014 cloud, which might harbor the precursor of a
young brown dwarf, although the distance to this cloud is not clear,  
and thus the newly detected object could be of higher mass \citep{young04, 
huard06}.

Jeff Linsky discussed the question of whether the EGGs in the Eagle Nebula are
early stages of low-mass star or brown dwarf formation and the possible
significance of the non-detection of X-rays from the EGGs. None of the EGGs are
detected in X-rays with luminosity upper limits below those of pre-main
sequence stars in the Orion Nebula Cluster. At the age of 2 Myrs, similar to 
the NGC6611 cluster whose bright stars evaporate the EGGs, young objects in 
the EGGs should show up in X-rays. However, the EGGs, and possible objects in
them, could be much younger -- their age is not really known.

\section{Spectral energy distributions and evidence for disks around 
proto-brown dwarfs}

The next two talks investigated the presence of disks around known proto-brown
dwarfs, addressing the question down to which masses of the central objects
disks could be detected, and if disks from very low-mass objects would show
very small disk masses as well, which could be attributed to a cutoff of the
disks during an ejection event.

Until recently, our detailed knowledge of substellar disks relied on a 
few case studies \citep*{apai02}. Alexander Scholz reported on 
a program that he and his collaborators carried out to characterise 
disks of brown dwarfs based on large object samples using the unique 
capabilities of $SPITZER$ combined with sensitive observations in the 
mm-range with the $MAMBOII$ bolometer array at the IRAM 30-m antenna. 
The analysis of the spectral energy distributions of brown dwarfs 
allows them to derive the properties and evolution of these disks, and 
additionally to constrain the efficiency and universality of planet 
formation.

The study at 1.3 mm in Taurus provided mid-infrared to mm spectral energy
distributions for 20 brown dwarfs, which allowed them to constrain disk 
masses and radii for these objects for the first time in a systematic way. 
Six out of the 20 brown dwarfs were actually detected at mm-wavelength, 
while for the others sensitive upper limits were placed. By combining 
these $IRAM$ with $SPITZER$ data, a minimum outer disc radius of 10 AU 
is necessary to interpret the spectral energy distributions in five cases. 
From these observations, there is no evidence for truncated disks due to an
ejection process early in the life of brown dwarfs, which implies that
most sub-stellar objects probably form in isolation. 

Alexander Scholz also reported on an ongoing study of brown dwarf disks in 
Upper Scorpius based on $SPITZER$ data. In this study, the infrared 
spectral energy distributions between the K-band and 24 $\mu$m of 26
brown dwarfs in Upper Scorpius, with ages of ~5 Myr, are investigated. 
While \citep{carpenter06} found no disks for solar mass stars 
in this region and at this age, about 37 percent of the brown dwarfs 
seem to have disks, although only a small fraction seems to be accreting. 
This would provide evidence that disks around low mass objects are longer
lived than disk around solar-type stars.

\begin{figure}[!ht]
\begin{center}
\includegraphics[width = 370pt, height = 370pt]{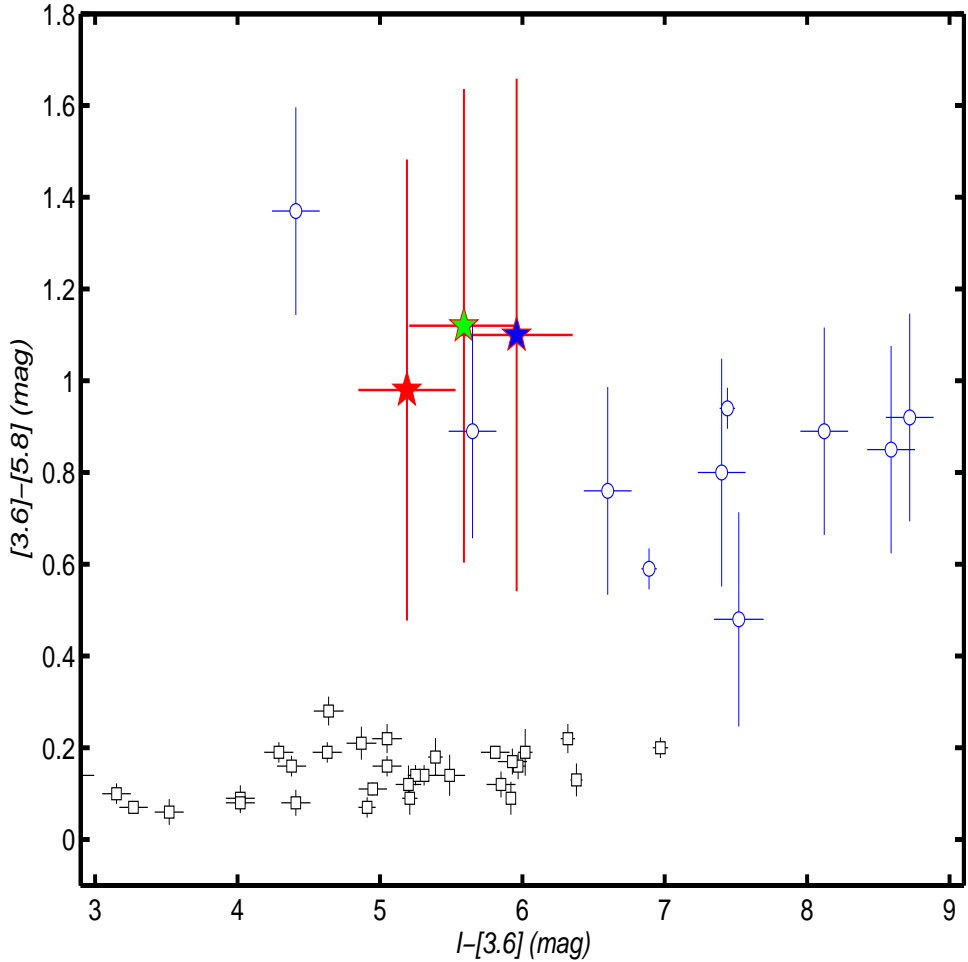}
\end{center}
\caption{$[3.6]-[5.8]$ vs. $I-[3.6]$ colour-colour diagram of three candidate
isolated planetary-mass objects in $\sigma$ Orionis with probable discs
(big filled stars) in comparison with field ultracool dwarfs with spectral
types in the range M6.5--L5.0 (open squares) and low-mass brown dwarfs and
IPMOs in Chamaeleon, Ophiuchus and Lupus with discs (open circles).}\label{fig1}
\end{figure}

Jos\'e Caballero presented a correlation of objects in the region around  
the O9.5V-type star sigma Ori A from the $DENIS$ and $2MASS$ catalogues and 
$IRAC$ data from the $SPITZER$ Space Telescope Data Archive. From the 
available photometry, he could produce spectral energy distributions in the 
IJHKs bands and at 3.6, 4.5, 5.8, and 8.0 $\mu$m for several thousand 
sources, and from infrared excesses infer the presence of disks surrounding 
a large fraction of members of the sigma Orionis cluster, from massive 
stars into the planetary mass regime. Especially interesting are three 
isolated planetary objects that show infrared excess at 5.8 $\mu$m, and 
thus belong to the lowest mass objects with disks.

\section{The mass function in the sub-stellar mass regime}

Finally, Morten Anderson presented a survey in the Mon R2 cluster in JHK
obtained with the $NICMOS$ camera on board the $HST$. The goal of this 
survey was to determine if the ratio of stars to brown dwarfs is universal, 
and what the slope of the initial mass function (IMF) for brown dwarfs and 
down below the deuterium burning limit might be. This study finds a 
star-to-brown dwarf ratio for Mon R2 that is similar to what other studies 
have found for other clusters \citep*{bejar01, moraux03}, and to 
within 2$\sigma$ is in agreement with the IMF in the solar neighborhood 
derived by \citep{chabrier02}. Morten Anderson therefore concluded that the 
IMF around and below the deuterium burning limit is falling or flat, but not
rising.

\section{Discussion}

Following these presentations, an open and lively discussion with the speakers
and the audience developed about the likely formation scenario for substellar
objects.

Most contributors took the view point that none of the recent observations
would favor the ejection scenario. Instead everything would hint at a
continuous formation process down to the lowest masses. In the absence of such
evidence it would certainly seem justified to assume a common mechanism for
the formation of all objects.

Very important open questions still remaining are how low in mass objects can
get, and how the mass function at the very lowest masses might look like. It
would also be very interesting to compare the core mass function with the mass
function in young clusters. At first glance, comparing the results presented
by Jane Greaves and Morten Anderson, it seems that the star formation
efficiency may (strongly) drop towards lower masses. This topic certainly
deserves closer investigation.

\acknowledgements 
The conveners of this splinter session would like to thank all speakers and
participants for their contributions and the lively discussion. It is a
pleasure to thank the organizers of the Cool Stars 14 conference, who made
this splinter session possible, and Patrick Lowrance for his help in running
it.
The work of J.E. was partially funded by Deutsche Forschungsgemeinschaft
(DFG), grant Ei 409/6.


\end{document}